\documentclass{article}

\PassOptionsToPackage{numbers}{natbib}

\usepackage[final]{neurips_2021_ml4ps}

\usepackage[utf8]{inputenc} 
\usepackage[T1]{fontenc}    
\usepackage{hyperref}       
\usepackage{url}            
\usepackage{booktabs}       
\usepackage{amsfonts}       
\usepackage{nicefrac}       
\usepackage{microtype}      
\usepackage{xcolor}         
\usepackage{graphicx}
\usepackage{enumitem}

\title{Self-supervised similarity search\\ for large scientific datasets}

\author{%
   George Stein \\
   Berkeley Center for Cosmological Physics \\
   Lawrence Berkeley National Laboratory\\
   Berkeley, CA 94720\\
   \texttt{gstein@berkeley.edu}\\
    \And
    Peter Harrington \\
   Lawrence Berkeley National Laboratory\\
   Berkeley, CA 94720\\
   \texttt{pharrington@lbl.gov}\\
    \And
    Jacqueline Blaum \\
    Department of Astronomy, University of California, Berkeley \\
    Lawrence Berkeley National Laboratory \\
    Berkeley, CA 94720 \\
    \texttt{jrblaum@berkeley.edu}\\
    \And
    Tomislav Medan \\
    Lawrence Berkeley National Laboratory\\
   Berkeley, CA 94720\\
    \texttt{tmedan@lbl.gov}
   \And
   Zarija Luki\'{c}\\
   Lawrence Berkeley National Laboratory\\
   Berkeley, CA 94720\\
   \texttt{zarija@lbl.gov}\\
}

\begin{document}

\maketitle

\begin{abstract}

We present the use of self-supervised learning to explore and exploit large unlabeled datasets. Focusing on 42 million galaxy images from the latest data release of the Dark Energy Spectroscopic Instrument (DESI) Legacy Imaging Surveys, we first train a self-supervised model to distill low-dimensional representations that are robust to symmetries, uncertainties, and noise in each image. We then use the representations to construct and publicly release an interactive semantic similarity search tool. We demonstrate how our tool can be used to rapidly discover rare objects given only a single example, increase the speed of crowd-sourcing campaigns, and construct and improve training sets for supervised applications. While we focus on images from sky surveys, the technique is straightforward to apply to any scientific dataset of any dimensionality. The similarity search web app can be found at \href{https://github.com/georgestein/galaxy\_search}{github.com/georgestein/galaxy\_search}.

\end{abstract}

\section{Introduction}

Scientific datasets continue to increase in both size and complexity, yet often lack high-quality labels that can be used to explore and understand the natural distribution of objects and features within them. As such, we require accessible automated tools to select sets of interesting objects and to assess the rarity of individual observations. One such important tool for exploration and data discovery is similarity search, which, given an input sample, rapidly returns semantically similar instances. Also known as query by example, content-based instance retrieval, or reverse instance search, this can be used for a number of scientific investigations including discovering rare objects given only a single example, flagging and identifying bad data, and rapidly constructing and improving training sets for supervised applications.

Quantifying the similarity between two data samples is difficult in the measurement space, as samples that contain the same scientific information can have significantly different data values. For example, rotating an image will likely drastically change individual pixel values, yet does not change the semantic information in that image (assuming rotational symmetry). Therefore, instance-based retrieval methods first extract a lower-dimensional representation from each data sample, and then construct a notion of distance or similarity in this feature space. While commonly used feature extraction methods such as t-SNE \citep{tsne}, UMAP \citep{umap}, pre-trained supervised networks \cite{SupervisedCNN}, or autoencoders and their variants \cite{HintonAE,VAE}, have been shown to extract relevant information for a number of tasks, they can also put focus on scientifically unimportant features or noise. In the case of pre-trained supervised networks, which rely on human-annotated labels, data samples can end up being grouped more by their predetermined (human-constructed) classes than purely by their semantic similarity. The key for extracting scientifically useful features is to explicitly inform the model of the symmetries, uncertainties, or instrumental noise in the dataset, so that the resulting features are robust to these perturbations.

\begin{figure}[t]
    \centering
    \includegraphics[width=1.0\textwidth]{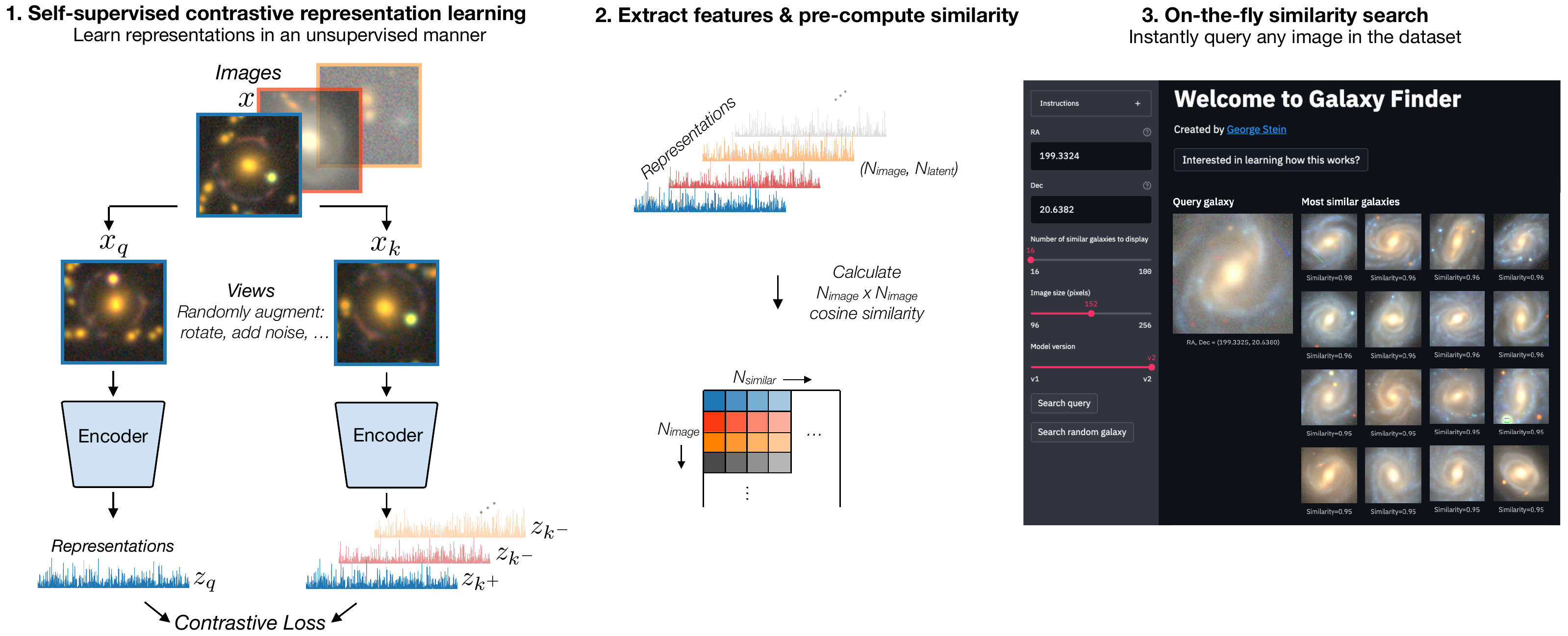}
    \caption{Contrastive self-supervised learning to extract representations which are robust to symmetries, uncertainties, and instrumental noise in the dataset (left), feature extraction and similarity computation (center), and our public interactive similarity search tool (right) -- \href{https://github.com/georgestein/galaxy\_search}{github.com/georgestein/galaxy\_search}.} 
    \label{fig:network}
\end{figure}

In this vein, contrastive self-supervised learning has achieved remarkable results in recent years, extracting features (from potentially vast uncurated datasets, as in \cite{SEER}) which have remarkable predictive power and semantic structure despite never using any supervision labels during training \cite{DINO,chen2020improved,BYOL,SimCLR}. 
Contrastive learning allows one to easily tailor the training process such that extracted features exclude undesired or uninformative properties, like symmetries or observational noise in a given dataset, by adjusting the data augmentations used. Thus, the features extracted from contrastive self-supervised models are an excellent choice for use in scientific similarity searches.
Here we present the use of self-supervised learning to distill low-dimensional representations from 42 million galaxy images from the latest data release of the Dark Energy Spectroscopic Instrument (DESI) Legacy Imaging Surveys, and then construct and publicly release a semantic similarity search tool that can instantly retrieve the most similar images to any given query. We show a schematic of our approach in Figure \ref{fig:network}. While we focus on an example from sky surveys, the technique we present can in principle be used on any scientific dataset of any dimensionality.

\section{Methods}
\paragraph{Data}
Our dataset consists of galaxy images from the latest data release of the DESI Legacy Imaging Surveys\footnote{\href{https://www.legacysurvey.org/}{legacysurvey.org/}}, Data Release 9 (DR9), released in January of 2021 \citep{DECaLS}. Our self-supervised approach allows us to learn from the entire dataset and does not require any labels or restrictive data cuts, so we construct a dataset of the brightest galaxies from the south region of DR9, as this region has the lowest level of noise. We focused on the 42,272,646 galaxies with a z-band magnitude $<$ 20, as galaxies dimmer than this threshold generally have an angular size that only covers a few pixels of the image. Centered on each galaxy we extracted a 152$\times$152 pixel cutout in the three optical bands (g, r, z) at the 0.262 arcsecond resolution of the telescope. Our network was trained on 96$\times$96 pixel crops, which is sufficient to cover the angular extent of all but the most near-by galaxies, but the $152^2$ pixel versions are required in order to perform the required set of data augmentations during training. This results in a dataset size of 10TB.

\paragraph{Self-supervised learning}
We closely follow \cite{Hayat_2021} in designing the architecture and training procedure for our self-supervised model, which is based on MoCov2 \cite{chen2020improved}. In this setting, the backbone of the model is a CNN encoder that takes an image $\bf{x}$ as input and produces a lower dimensional representation $\bf{z}$. The encoder learns to make meaningful representations by associating augmented views of the same image as similar, and views of different images as dissimilar, via a contrastive loss function. We use the same ResNet50 network and training hyperparameters as \cite{Hayat_2021}, but increase the queue length to $K=262,144$ to accommodate our larger training set. We choose the following set of augmentations, applying each of them in succession to images during pre-training with the order listed below \citep[see][for motivation]{stein}:

\begin{itemize}[leftmargin=*]
     \itemsep0em 

\item \textit{Galactic extinction:} We de-redden the image according to its tabulated SFD $E(B-V)$ value \citep{SFD1998Dust}, then randomly sample a new $E(B-V)$ value from a log-normal distribution fit to the dataset.

\item \textit{Rotation/Orientation:} We randomly flip the image across each axis with 50\% probability, then rotate by a random angle sampled from $\mathcal{U} (0, 2\pi)$.

\item \textit{Size Scaling:} We randomly re-size the image to between 90\% and 110\% of its original size by rescaling with bilinear interpolation.

\item \textit{Point Spread Function (PSF) blur:} We sample a Gaussian blur for each channel by sampling from log-normal fits to the PSF distribution of the data.

\item \textit{Jitter and crop:} We translate the image center along each respective axis by $\mathcal{U}(-7,7)$ pixels, before cropping out the central 96$\times$96 pixels.

\item \textit{Gaussian noise:} We sample a Gaussian noise level from log-normal distributions fit to the noise distribution in each filter channel. 

\end{itemize}

With these augmentations, we perform self-supervised pretraining on a curated subset of 3.5 million galaxies sampled uniformly by $z$-band magnitude. Including all objects was deemed to be not worth the order of magnitude more computation time, as the majority of the fainter galaxies have a spatial size in the image spanning no more than a few pixels. We are able to train the model in just 8 hours on 8 NVIDIA Tesla V100 GPUs. Then, we apply the CNN encoder to all images in our dataset, extracting a representation vector $\mathbf{z}$ of length 2048 for each galaxy.

\paragraph{Similarity Search} Self-supervised pre-training provides a robust measure of semantic similarity between any set of images. The representations extracted from each image in the dataset retain the overall information such as the size and relative orientation of the galaxies, the number of sources, the clustering, and the color, while removing the noise and symmetries described by the data augmentations \citep[for a visualization of how galaxy properties are organized in the representation space of a self-supervised model, see][]{Hayat_2021}. 

The contrastive loss in self-supervised pre-training encourages semantically similar images to have similar representations, so we can easily define a similarity metric in representation space (which is a lower dimensionality than the input images). We choose to measure similarity between any two images $\bf{x_i}$ and $\bf{x_j}$ by the cosine similarity (normalized scalar product) of their representation vectors $\bf{z_i}$ and $\bf{z_j}$:
\begin{equation}
    \mathrm{similarity}(\bf{z_i},\bf{z_j}) = \bf{z_i} \cdot \bf{z_j}/ (\| \bf{z_i} \| \, \| \bf{z_j} \| ).
\end{equation}
To speed up similarity search we pre-calculate the 999 most similar images to each of the 42 million images, and save the resulting ``similarity array'' (size $\sim$42M$\times$1,000) for rapid querying. Computations are performed on 8 GPUs using Facebook AI Research's Faiss library for efficient similarity search.\footnote{\href{https://github.com/facebookresearch/faiss}{github.com/facebookresearch/faiss}. While we calculate similarity without employing any of the compression techniques available in Faiss we found that it still outperformed its NumPy equivalent on CPU, and benefited from an additional speed up on GPU} To avoid unnecessary computations we did not compute the full $N\times N$ similarity matrix ($N=42$ million), but for each source limited the search to sources within a magnitude range of 0.5 in r-band, as we found that this resulted in nearly equivalent results as the full search due to the model naturally learning that brighter galaxies do not have significant similarity with dimmer ones.

\section{Results}

\begin{figure}[t]
    \centering
    \includegraphics[width=0.49\textwidth, trim={0.18cm 8.475cm 0.25cm 8.475cm}, clip]{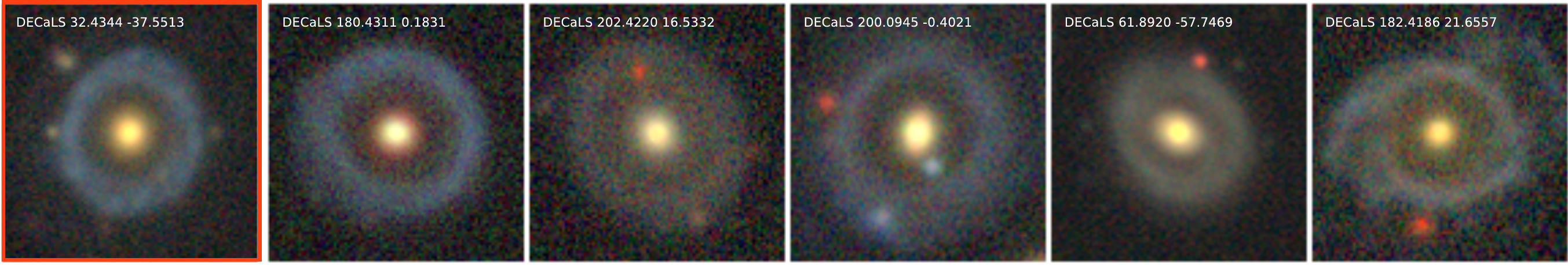}
    \includegraphics[width=0.49\textwidth, trim={0.18cm 8.475cm 0.25cm 8.475cm}, clip]{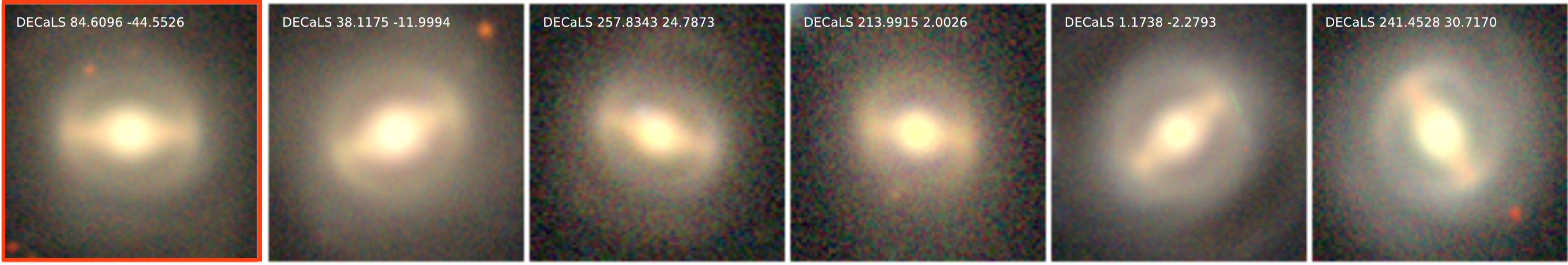}
    \includegraphics[width=0.49\textwidth, trim={0.18cm 8.475cm 0.25cm 8.475cm}, clip]{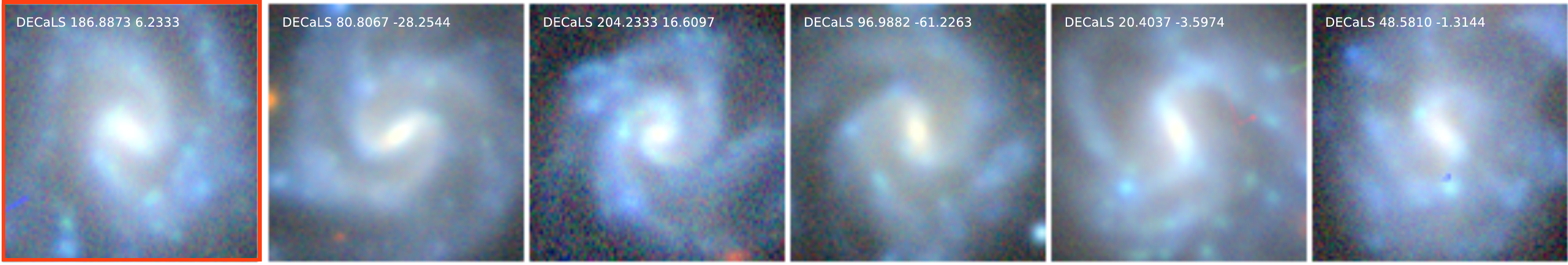}
    \includegraphics[width=0.49\textwidth, trim={0.18cm 8.475cm 0.25cm 8.475cm}, clip]{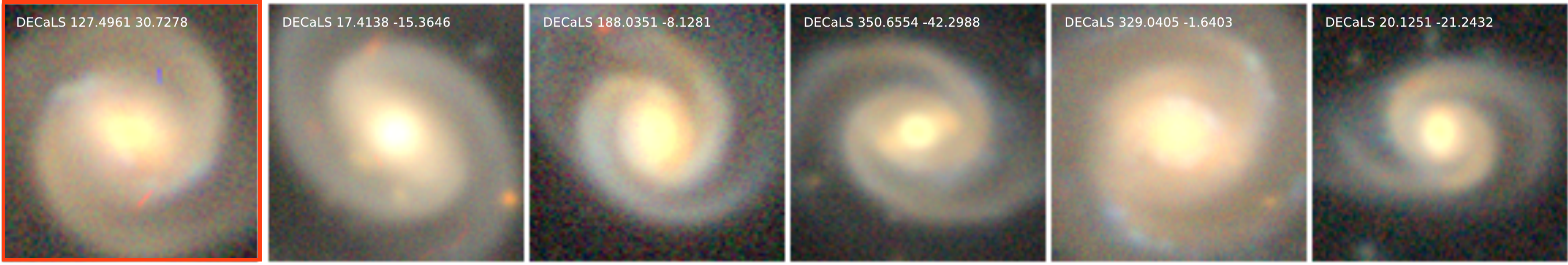}
    \includegraphics[width=0.49\textwidth, trim={0.18cm 8.475cm 0.25cm 8.475cm}, clip]{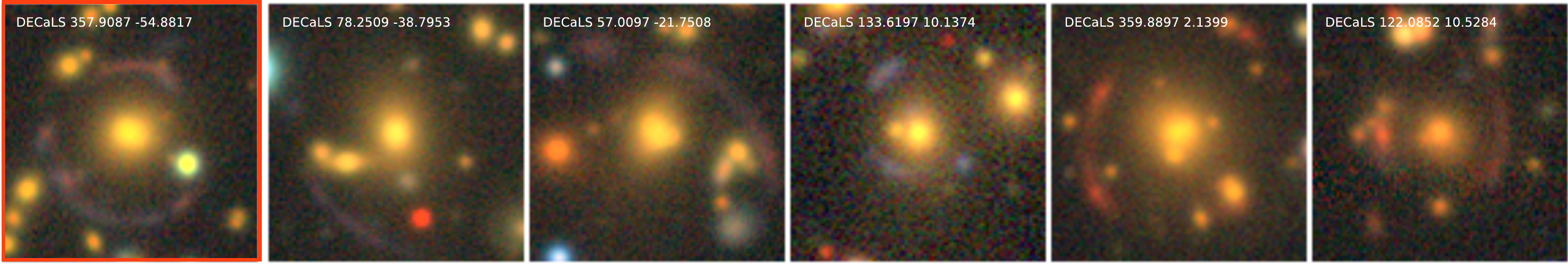}
    \includegraphics[width=0.49\textwidth, trim={0.18cm 8.475cm 0.25cm 8.475cm}, clip]{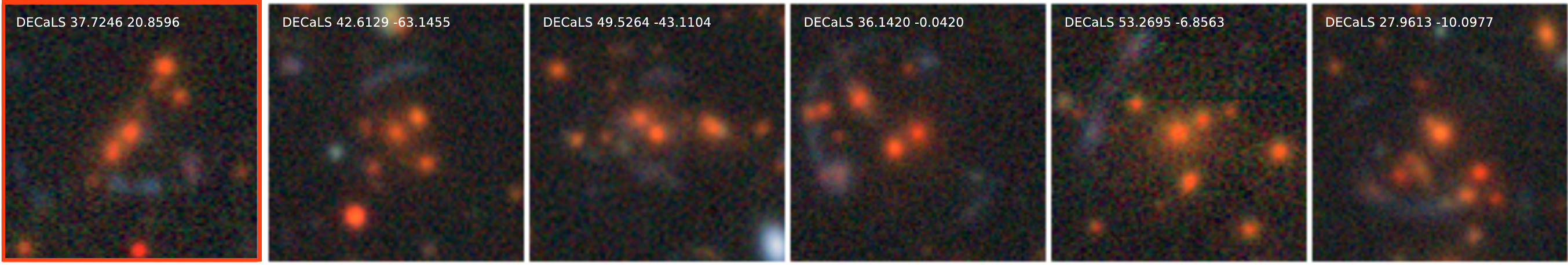}
    \includegraphics[width=0.49\textwidth, trim={0.18cm 8.475cm 0.25cm 8.475cm}, clip]{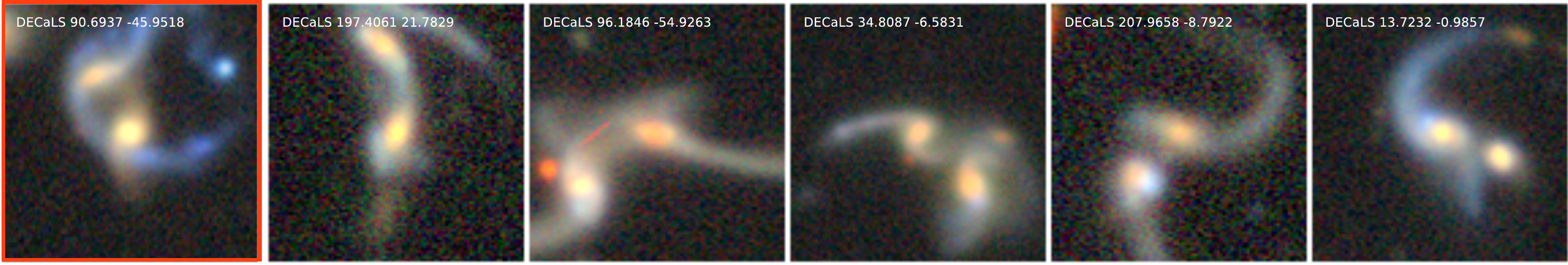}
    \includegraphics[width=0.49\textwidth, trim={0.18cm 8.475cm 0.25cm 8.475cm}, clip]{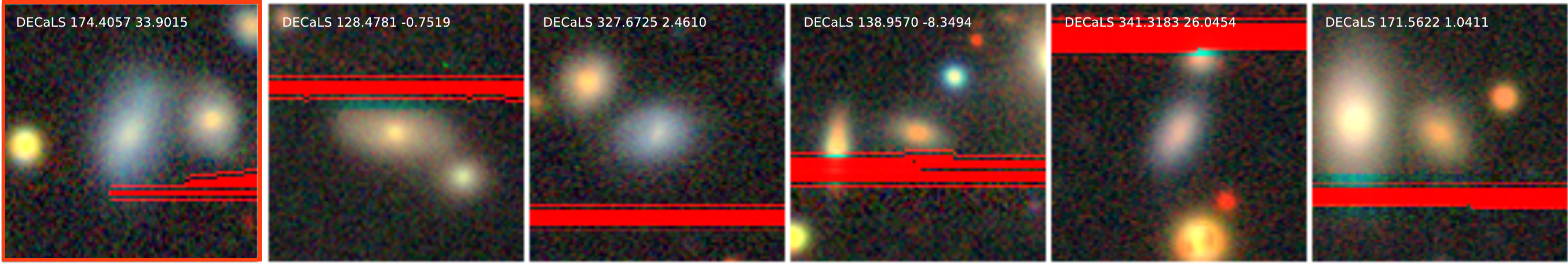}
    \caption{Query examples (left-most panels, red outline) and a sample of five similar galaxies returned through self-supervised similarity search, showing ring galaxies (top row), spiral galaxies (second row), strong gravitational lenses (third row) \citep{stein}, and galaxy mergers and images corrupted by nearby bright stars (bottom row).} 
    \label{fig:similarity_search}
\end{figure}

We first experimented with which representation layer to use for optimal similarity computation. The layer before the projection head, rather than the final layer where the contrastive loss is computed, has been shown to improve linear classification results on ImageNet by >10\% \citep{SimCLR}, but this does not necessarily directly transfer to an improved similarity search. We construct separate similarity search models using both sets of representations (2048 dimensional and 128 dimensional representations, respectively), and augment a large random sample of images with the same transformations used during training. Performing similarity search with the original images as a query, we find that the augmented versions are nearly always returned as the most similar sample to the original image for both models, with little to no quantitative difference. While perhaps unsurprising given the clear semantic similarity of query examples and the most similar images returned, this test provides evidence that both representation layers of our self-supervised model are robust to the input perturbations, as designed. Whether this is a result of our augmentations being an incomplete description of the full uncertainty in the dataset is left for future work. From visual inspections of a number of similarity searches we conclude that the most similar images returned by the pre-projection (2048-dimensional) representations marginally outperform those from the final post-projection representations. 

In Figure~\ref{fig:similarity_search} we provide a few example searches using astronomical objects of interest, and show similar galaxies hand-selected from the top $\sim 10$ returned. As gravitational lenses are exceptionally rare, for these two panels we instead examine the top few hundred examples returned and select the most interesting. We find remarkable results for a large diversity of objects -- the semantic similarity as measured by our self-supervised model goes far beyond what is possible by hand-selected color or magnitude cuts or hand-constructed filters. 
It is clear that the galaxies returned for each query do not simply have similar orientations or noise levels -- rather, the notion of similarity in the self-supervised representations extends far beyond such basic characteristics, and reflects the rich set of object categories in the data.

For lack of space we highlight only two examples:
1.) Strong gravitational lens finding poses a significant challenge, as they are amongst the rarest objects in the universe ($\sim$1 in $\mathcal{O}(10^4)$ \citep{Collett}). Finding them requires large visual inspection efforts with standard supervised learning techniques due to limitations in the small number of labels available for training \citep[see][and references within]{strong_lens_challenge}, yet here we queried for a single serendipitously discovered lens \citep{strong_lens} (third row, on left) and instantly discovered many additional examples. In \citep{stein} we utilize similarity search alongside supervised classification to discover $\sim$1,200 new strong lenses. 2.) Ongoing crowd sourcing campaigns to discover and classify objects such as ring-galaxies\footnote{see the ongoing crowd-sourced ring galaxy campaign \href{https://blog.galaxyzoo.org/2021/09/23/new-galaxy-zoo-mobile-challenge-ring-galaxies/}{blog.galaxyzoo.org/2021/09/23/new-galaxy-zoo-mobile-challenge-ring-galaxies/}} can be massively sped up through self-supervised similarity search by pre-selecting unlabelled samples based on minimal similarity searches, and then inspecting them in parallel.

Similarity search on self-supervised representations may focus on any features that are not explicitly excluded by the data augmentations. For example, an input query to our search tool that contains bright foreground stars or galaxies can have the ``similarity focus'' drawn away from object of interest, and will instead return other images with similar foreground features. In this way this technique is limited by the augmentations chosen during training. A potential workaround is to include such foreground effects as data augmentations in self-supervised pre-training --- for example by randomly adding bright foreground objects to a given view of the original image --- which would incentivize the representations to be less susceptible to these perturbations. We leave this for future investigation. Additionally, for this work we chose a minimal set of task-agnostic augmentations to use during training, such that the representations are directly useful for a broad number of survey science tasks. If the tool was instead desired for more specific tasks, for example an emphasis on color over shape, the strength of the augmentations, or the inclusion of additional ones, can help shape the similarity criteria. 

\section{Conclusion}
We have demonstrated that a self-supervised learning model can be used to extract information-rich representations from unlabelled data that are robust to symmetries, uncertainties, or instrumental noise in the dataset. These representations provide a direct quantification of similarity between any two data samples, an extremely useful ability given that most real-world scientific datasets have no objective measure of ``true'' similarity between samples. As the model was trained only to distinguish the natural distribution of features within the data sample, it is not limited to identification of a specific class of object, but can be used to facilitate discovery of any object(s) that exist in the dataset.


\section{Broader impact}

Generalized self-supervised models greatly reduce the barrier to entry for working with modern datasets, and open up a number of collaborative avenues that were previously unavailable. Rather than each team working alone to perform classification or regression tasks on large scientific datasets -- applying for time on large GPU computing systems, downloading massive datasets, learning to train models in parallel, etc. -- the initial model training can be undertaken separately, and the fully trained model and image/representation pairs can be shared to allow for rapid scientific investigations that are not hindered by computational complexity or resource allocation. This creates the potential to further democratise scientific investigations on large datasets to institutions that do not have the same computational accessibility as tier-one research institutes or large technology companies. 

\section{Acknowledgements}
We would like to thank Rongpu Zhou for his significant help on using the DESI Legacy Survey data, Dustin Lang for providing access to the image-cutout service at NERSC, and Md Abul Hayat and Mustafa Mustafa for their pioneering efforts on self-supervised learning for sky surveys.

This research used resources of the National Energy Research Scientific Computing Center (NERSC), a U.S. Department of Energy Office of Science User Facility located at Lawrence Berkeley National Laboratory, operated under Contract No. DE-AC02-05CH11231. This research also used resources of the Argonne Leadership Computing Facility, which is a DOE Office of Science User Facility supported under Contract DE-AC02-06CH11357. G.S., J.B., T.M., and Z.L.~were partially supported by the DOE's Office of Advanced Scientific Computing Research and Office of High Energy Physics through the Scientific Discovery through Advanced Computing (SciDAC) program.

\bibliographystyle{abbrvnat}
\bibliography{ref}

\end{document}